\documentclass[seceq]{ptptex}

\usepackage{graphicx}
\notypesetlogo                       %comment in if to eliminate PTPTeX 
\def\oneone{\rlap 1\mkern4mu{\rm l}}
\def\coeff#1#2{\relax{\textstyle {#1 \over #2}}\displaystyle}

\title{%        %You can use \\ for explicit line-break
Microstate Geometries  and 
Entropy Enhancement%
}

\author{%       %Use \scshape  for the family name
Nicholas P. \textsc{Warner}%
}

\inst{%     %Affiliation, neglected when [addenda] or [errata]
\centerline{Department of Physics and Astronomy}
\centerline{University of Southern California}
\centerline{Los Angeles, CA 90089-0484, USA.}
}

\abst{%         %this abstract is neglected when [addenda] or [errata]
In string theory, black-hole backgrounds are far from unique and there are large families of completely smooth, horizonless geometries with the same structure as a black hole from infinity down to the neighborhood of the black-hole horizon.   These microstate geometries cap off in foam of topological bubbles.  I review some of the recent progress in constructing these smooth horizonless geometries and discuss some of the physical implications.  }

\begin{document}

\maketitle

\section{Introduction}

String theory is, without doubt, one of the most interesting  theoretical ideas to emerge in particle physics in the last thirty years and thus  has attracted many researchers and great deal of attention.  While there has been vast progress in developing string theory itself, there has also been a growing concern about whether string theory will provide tangible progress in physics and ultimately lead to testable predictions.  

Finding ways to probe Planck-scale physics is a huge (but probably not impossible) challenge.  On the other hand, there are also  more indirect ways to test string theory and determine the extent of its contribution to science.  First, it is beyond doubt that the spin-offs to other fields, ranging from pure mathematics to the detailed computation of Feynman diagrams in field theory,  have been remarkable if not revolutionary.  More directly, the last twelve years has seen a growing body of work that begins to address some of the deep, long-standing problems  in black-hole physics and that makes connections to experimental physics.   One of the most powerful tools to emerge is the idea of holographic field theories and this has not only deepened our understanding of strongly coupled field theory but is also giving interesting insights into new experimental data about  quark-gluon plasmas and non-relativistic conformal field theories.   This diversity of application and the range and breadth of the talks at this meeting not only highlights the health and vibrancy of string theory research but also shows that string theory is addressing important open problems in physics.
  
It is also remarkable that Prof. Eguchi has made major contributions to the remarkably diverse areas of theoretical physics covered by this conference.  This is a testament to his health and vibrancy as a physicist and his judgment in seeing what is, or will be,  important.  Prof. Eguchi has also influenced my work significantly ever since I was student, starting with  his extremely useful {\it Physics Reports} \cite{Eguchi:1980jx}, progressing through his work on gravity, conformal field theory, integrable models, topological models and finally supersymmetric field theories.  Indeed, as I will discuss, Prof. Eguchi's work on   gravitational instantons \cite{Eguchi:1978gw, Eguchi:1979yx} is once again finding application in the microstate structure of black holes.

This talk is primarily about addressing a problem that has vexed gravitational physics for over thirty years:  The black-hole microstate problem.  From thermodynamic arguments in general relativity, we know that a black hole has a vast entropy and yet black-hole uniqueness in general relativity states that given the bulk thermodynamic quantities, like mass, charge and angular momentum, there is a unique solution and so there are no microstates that can account for this entropy.  It is  popular to state that the cosmological constant is the largest problem in theoretical physics, however for the black hole in the middle of our galaxy (with a mass of about $4 \times 10^6 M_\odot$) the entropy is about $2 \times 10^{90}$, and remembering that the entropy is the {\it lograrithm} of the number of microstates, the counting of microstates  is off by a factor of $e^{10^{90}}$.

String theory has already made significant progress in addressing this issue.  One may describe a supersymmetric, BPS black hole in terms of D-branes and in the weak coupling limit  one can count the number of different BPS configurations with the same asymptotic charges \cite{Sen:1995in, Strominger:1996sh} and the result matches with the thermodynamic entropy associated with the horizon area.   While this work is remarkable, it makes very significant restrictions on the black hole:  It must be supersymmetric and the analysis was only done for weak coupling.  The former restriction makes the black hole very simple and rather special,  essentially because it has vanishing Hawking temperature, while the latter makes the system look nothing like a black hole because gravitational back-reaction has been turned off.  However, the number of BPS states is expected to be preserved under deformations and so turning on the gravitational coupling will not change the counting of microstates.  This expectation is  born out by the principles of holography and the AdS/CFT correspondence, where the gravitational back-reaction plays an essential role. While the state counting does not change, the form of the microstates can, and will, change radically and so the obvious question is what do these microstates look like at finite string coupling?  

Obviously one would also like to address the microstate structure of a non-BPS black hole, or better,  simply a Schwarzschild or Kerr black hole.  While this is beyond our reach at present,  I will make some remarks about near-BPS black holes at the end of this talk.  

The parallels between the history of the weak coupling analysis in the last decade and the more recent finite-coupling analysis are quite close.  The weak coupling analysis of microstates was first performed by Sen \cite{Sen:1995in} for the two-charge system in five dimensions and, while extremely interesting and very suggestive, it was not an instant revolution:  The problem was that the two-charge system does not have a macroscopic horizon and so one must appeal to a ``stretched'' or ``effective'' horizon.  The revolution came when Strominger and Vafa \cite{Strominger:1996sh} performed the computation   for the three-charge system, which has a truly macroscopic horizon and so one has faith in the underlying supergravity description.   At finite coupling, the computations are much more difficult but a remarkable body of work has been done by Mathur and collaborators (see Ref. \citen{Mathur:2005zp} for a review)   for the two-charge system and a very interesting and fairly compelling case has been made.  On the other hand, the semi-classical configurations that are supposed to account for the black-hole microstates of the two-charge system have a scale that is dangerously close to the string scale and typical states that contribute to the entropy are almost certainly string scale.  Thus there is a range of opinions about the work on the two-charge system but there is complete agreement that the  ideas pioneered by Mathur must be examined for three-charge systems where black-hole horizons are macroscopic.  There has been much progress on this in the last four years and my intention here is to review some of that work.

The basic core question is that given a set of boundary conditions that would define a unique black hole within general relativity and electromagnetism, what are the possible geometries that fit these boundary conditions in string theory or M theory?   In particular,  a {\it Microstate Geometry} is defined to be any  {\it completely smooth, horizonless} solution that matches the   boundary conditions of a given black hole.  One of the surprises of the last three years has been that there are a vast number of such geometries and a very rich underlying structure that matches very nicely with the dual holographic field theory description.    It remains to be seen if these can provide a semi-classical accounting of the entropy but, as I will discuss,  there is a reasonable chance that they might.

Independent of the microstate counting issue, the study of microstate geometries is extremely interesting.  It is obviously important to understand the failure of black-hole uniqueness in string theory and to classify all possible solutions with the same asymptotics as a given black hole.  More generally, given that there are vast number of smooth solutions that lie within the validity of the supergravity approximation and have the same structure at infinity as a black hole, one must determine their role within black-hole physics.  Indeed, within string theory, the appearance of a black-hole singularity (at least for BPS solutions) is very much an artifact of imposing spherical symmetry: once the symmetry restriction is removed one has a vast number of smooth solutions.   Put simply, if one is presented with a choice between a smooth solution and a singular solution for a given set of boundary conditions then the burden of proof lies with the physicist who elects to say that the singular solution is a more accurate representation of reality than the regular solution.  Thus, even if there are not enough microstate geometries within the supergravity approximation to account for the entropy, the study of microstate geometries will probably change the way in which we describe the space-time in the interior of a black hole.

\section{Three-charge solutions in five dimensions}

Supersymmetric black holes with macroscopic horizons, and their corresponding microstate geometries, are most easily constructed as three-charge geometries in five dimensions using M theory.  The results can then easily be reduced to four dimensions and converted to any string duality frame.  One can use any Calabi-Yau manifold, but the essential features can be obtained via a   compactification on $T^6$.
The metric in eleven dimensions is therefore taken to be:
\begin{eqnarray}
 ds_{11}^2  = ds_5^2 & + &    \left(Z_2 Z_3  Z_1^{-2}  \right)^{1\over 3}
 (dx_5^2+dx_6^2) \nonumber \\
 & + & \left( Z_1 Z_3  Z_2^{-2} \right)^{1\over 3} (dx_7^2+dx_8^2)    +
  \left(Z_1 Z_2  Z_3^{-2} \right)^{1\over 3} (dx_9^2+dx_{10}^2) \,,
\label{elevenmetric}
\end{eqnarray}
for some functions $Z_I$.  The  five-dimensional space-time metric must have the form:
\begin{equation}
ds_5^2 ~\equiv~ - \left( Z_1 Z_2  Z_3 \right)^{-{2\over 3}}  (dt+k)^2 +
\left( Z_1 Z_2 Z_3\right)^{1\over 3} \, h_{\mu \nu}dy^\mu dy^\nu \,,
\label{fivemetric}
\end{equation}
for some one-form field, $k$, defined upon the spatial section of this
metric.  To describe a five-dimensional black hole, the metric must be asymptotic to flat 
$\Bbb{R}^{4,1} \times T^6$, and so we require the warp factors, $Z_I$, to limit to constants at infinity
and the four-metric
\begin{equation}
ds_4^2 ~\equiv~ h_{\mu \nu}dy^\mu dy^\nu \,,
\label{fourmetric}
\end{equation}
must limit to the flat, Riemannian (positive definite) metric on $\Bbb{R}^4$ at spatial infinity.  I will refer to the four-manifold with this metric as the base and denote it by ${\cal B}$.

The three $U(1)$ gauge fields, $A^{(I)}$, in five dimensions come from  the three-form 
Maxwell potential via the Ansatz:
\begin{equation}
C^{(3)}  = A^{(1)} \wedge dx_5 \wedge dx_6 ~+~  A^{(2)}   \wedge
dx_7 \wedge dx_8 ~+~ A^{(3)}  \wedge dx_9 \wedge dx_{10}  \,.
\label{Cfield:ring}
\end{equation}
To fix the normalization of these fields, I will take $Z_I  \to 1$ at infinity, thereby fixing the asymptotic volumes of the $T^2$ factors of the compactification.

To find the microstate geometries corresponding to a given supersymmetric black hole one assumes that these geometries possess the same supersymmetries as the corresponding black hole\footnote{It is conceivable that there might be microstate geometries whose supersymmetries undergo some form of dielectric polarization in the deep interior, but I will not consider that here.}.  Thus the supersymmetry is characterized by the projectors associated with M2 branes wrapping each $T^2$:
\begin{equation}
\big(\oneone ~-~  \Gamma^{056}) \, \epsilon ~=~  \big(\oneone ~-~
\Gamma^{078}) \, \epsilon ~=~  \big(\oneone ~-~  \Gamma^{09\,10}) \, \epsilon ~=~  0  \,.
\label{susyproj}
\end{equation}
Since the product of all the gamma-matrices is the identity matrix, this implies
\begin{equation}
\big(\oneone ~-~  \Gamma^{1234}) \, \epsilon ~=~   0  \,,
\label{fourhelicity}
\end{equation}
which means four-metric must be ``half-flat.''   Equivalently,  the Riemann tensor of (\ref{fourmetric}) must be self-dual,  and hence  have $SU(2)$ holonomy.  Either way, this means that  the metric (\ref{fourmetric})  on the base, ${\cal B}$, must be hyper-K\"ahler.

Such metrics were studied a long time ago as ``Gravitational Instantons,'' and one of the first non-trivial examples was found by Prof. Eguchi \cite{Eguchi:1978gw, Eguchi:1979yx}.  Much to the disappointment of the Euclidean quantum gravity program, it was shown that the only non-trivial examples of gravitational instantons were ALE spaces and that the only smooth, Riemannian, hyper-K\"ahler, four-dimensional metric that is asymptotic to $\Bbb{R}^4$ is, in fact, globally $\Bbb{R}^4$.  It would therefore appear that we must limit the base metric to $\Bbb{R}^4$ here, but, as we will see, there are vastly more possibilities that enable one to apply many of the ideas of gravitational instantons and space-time foam to the microstate structure of black holes.

If one introduces the ``dipole field strengths,''  $\Theta^{(I)}$:
\begin{equation}
\Theta^{(I)} ~\equiv~  d A^{(I)} + d\big(  Z_I^{-1} \, (dt +k) \big)  \,,
\label{Thetadefn}
\end{equation}
then the most general supersymmetric configuration is obtained by solving a {\it linear system}  \cite{Bena:2004de} of  {\it BPS equations}:
\begin{eqnarray}
 \Theta^{(I)}  &~=~&  \star_4 \, \Theta^{(I)} \label{BPSeqn:1} \,, \\
 \nabla^2  Z_I &~=~&  {1 \over 2  }  C_{IJK} \star_4 (\Theta^{(J)} \wedge
\Theta^{(K)})  \label{BPSeqn:2} \,, \\
 dk ~+~  \star_4 dk &~=~&  Z_I \,  \Theta^{(I)}\,,
\label{BPSeqn:3}
\end{eqnarray}
where $\star_4$ is the Hodge dual taken with respect to the
four-dimensional metric $h_{\mu \nu}$, and the structure
constants\footnote{If the $T^6$ compactification manifold is replaced
by a more general Calabi-Yau manifold, then the $C_{IJK}$ change
 accordingly.}  are given by $C_{IJK} ~\equiv~ |\epsilon_{IJK}|$. 

The last two equations allow the addition of homogeneous solutions.  In (\ref{BPSeqn:2}), adding such additional sources simply amounts to putting   BPS black holes into the background.  In (\ref{BPSeqn:3}) the homogeneous solution must be tuned extremely carefully so as to remove closed time-like curves (CTC's) from the solution.

There are essentially two options for the sources for $\Theta^{(I)}$.  One can choose singular line sources on the base metric or, if there is non-trivial cohomology in $H^2({\cal B}, \Bbb{R})$,  the $\Theta^{(I)}$ can be chosen to be linear combinations of the cohomological fluxes.  The singular line sources correspond to M5 branes wrapping a $T^4$ of the compactification and following a  profile in the base, ${\cal B}$, and such solutions can be used to make general three-charge black rings and supertubes \cite{Bena:2004de, Elvang:2004ds, Gauntlett:2004qy}.  To obtain smooth, normalizable field strengths,  $\Theta^{(I)}$, there must be non-trivial compact  $2$-cycles and this is only possible if the base is actually a non-trivial  hyper-K\"ahler manifold.  Thus finding smooth microstate geometries is contingent upon getting around the restriction to an $\Bbb{R}^4$ base.

\section{The geometric transition}

\subsection{The structure of the transitioned geometry}

One can get further insight into how string theory naturally leads to smooth microstate geometries that involve a foam of non-trivial $2$-cycles on the base by considering how to resolve the singular geometry of supertubes.

\begin{figure}
\centerline{\includegraphics[width=13.5 cm,height=3 cm] {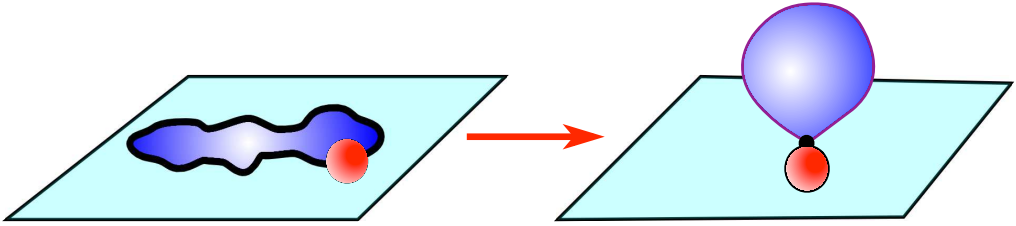}}
\caption{The diagram on the left shows the supertube profile in the unphysical base metric.  The profile is spanned by  a shaded disk and is surrounded by a Gaussian sphere.  In the physical metric (on the right) this supertube has zero length and the configuration transitions to two non-trivial homology cycles intersecting at a point where the supertube used to be. }
\label{fig:NPW1}
\end{figure}

In Mathur's original ``fuzzball proposal'' for two-charge systems, the semi-classical objects that account for the entropy can be described in terms of two-charge supertubes with arbitrary profiles.  In the D1-D5 duality frame the supergravity solutions are completely regular and thus yield microstate geometries.  One can generalize these solutions to the three-charge system but they become singular near the supertube.  Part of the singular nature of these solutions is that they have a null orbifold singularity in that the light cone is tangent to the closed curve that defines the supertube.   The supertube profile therefore appears to have finite length in the base metric (\ref{fourmetric}) but actually has vanishing length in the physical metric (\ref{fivemetric}).  As a result,  any disk that spans the supertube profile is actually pinched off into an $S^2$.  This is depicted in Figure \ref{fig:NPW1}.
%
%%%%%%%%%%%%%%   Figure 1   %%%%%%%%%%%%%%%
%
%
%%%%%%%%%%%%%%   End Figure 1   %%%%%%%%%%%%% 
%

The other important difference between the base metric and the physical metric is that a Gaussian $S^2$ that surrounds the supertube can have arbitrarily small size on the base metric, but has a finite size (set by the charges) in the physical metric. Thus, we are looking for a geometric transition that replaces the supertube in a flat background with two homology $S^2$'s in a curved background (see Figure \ref{fig:NPW1}).   Such geometric transitions are, by now, very familiar in string theory backgrounds and they involve the replacement of singular brane sources  by quantized fluxes   supported by non-trivial topology generated by the pinching effects of branes. \cite{Gopakumar:1998ki,Vafa:2000wi, Lin:2004nb, Cachazo:2001jy}.  Thus, in the geometric transition we seek,  the M5 brane fluxes, $\Theta^{(I)}$,  sourced by the singular supertube profile are to be replaced by smooth cohomological fluxes that are dual to the new homology spheres. 

The guiding principle in finding the transitioned geometry is that one should start from a hyper-K\"ahler base and seek a completely smooth, five-dimensional physical metric with no CTC's.  There must be no singular sources anywhere; the only sources must arise from cohomological fluxes dual to the $2$-cycles in the hyper-K\"ahler base.

\subsection{Gibbons-Hawking metrics}

To find examples of the geometric transitions that provide microstate geometries one needs to start with a simple, explicit  class  of hyper-K\"ahler four-manifolds. Probably the simplest such class is the family of Gibbons-Hawking metrics and these may be written as  $U(1)$ fibrations over a flat $\Bbb{R}^3$ base:
\begin{equation}
h_{\mu\nu}dx^\mu dx^\nu ~=~ V^{-1} \, \big( d\psi + \vec{A} \cdot d\vec{y}\big)^2  ~+~
 V\,  d\vec y \cdot d \vec y \,.
\label{GHmetric}
\end{equation}
%\
The function, $V$, is harmonic on the flat  $\Bbb{R}^3$ while the connection, $A = \vec A \cdot d\vec{y} $, is related to $V$ via
\begin{equation}
\vec \nabla \times \vec A ~=~ \vec \nabla V\,.
\label{AVreln}
\end{equation}
I should stress that we focus on this class of metrics simply for computational convenience and one can, in principle, study broader classes of hyper-K\"ahler metrics in four dimensions (see, for example, Ref. \citen{Bena:2007ju}).

In the standard form of the Gibbons-Hawking (GH) metrics one takes $V$ to have a finite set  of isolated sources.   That is, let  $\vec{y}^{(j)}$ be  the positions of the source points in the $\Bbb{R}^3$  and let  $r_j \equiv |\vec{y}-\vec{y}^{(j)}|$.  Then one takes:
\begin{equation}
 V = \varepsilon_0 ~+~ \sum_{j=1}^N \,  {q_j  \over r_j} \,,
\label{Vform}
\end{equation}
where one usually requires $q_j \ge 0$ to ensure that the metric is Riemannian.   We will later relax this restriction. 

There appear to be singularities in the metric at $r_j =0$, however, if one changes to polar coordinates centered at $r_j =0$ with radial coordinate to $\rho = 2 \sqrt{
|\vec{y}-\vec{y}^{(j)}|}$, then the metric is locally of the form:
\begin{equation}
ds_4^2 ~\sim~ d \rho^2 ~+~ \rho^2 \, d \Omega_3^2 \,,
\label{asympmet}
\end{equation}
where $d \Omega_3^2$ is the standard metric on $S^3/\Bbb{Z}_{|q_j|}$.
In particular, this means that one must have $q_j \in\Bbb{Z}$ and if $|q_j| = 1$ then the space looks locally like $\Bbb{R}^4$.  If $|q_j| \ne 1$ then there is an orbifold singularity, but since this is benign in string theory, we will view such backgrounds as regular.  

Similarly, at infinity one has 
\begin{equation}
V ~\sim~  \varepsilon_0 ~+~ {q_0 \over r} \,,
\label{asympV}
\end{equation}
where $r = |\vec y|$.    For the $U(1)$ fiber to decompactify and become large in four dimensions\footnote{By keeping  $\varepsilon_0  \ne 0$ one can generalize the results here to microstate geometries in four dimensions \cite{Gaiotto:2005gf,Gaiotto:2005xt, Elvang:2005sa, Bena:2005ni}.} one must take $ \varepsilon_0  =0$ and then the metric (\ref{GHmetric}) is asymptotic to the flat metric on $\Bbb{R}^4/\Bbb{Z}_{|q_0|}$, where 
\begin{equation}
q_0 ~\equiv~ \sum_{j=1}^N \, q_j \,.
\label{qzerodefn}
\end{equation}
For a positive definite, smooth base metric on ${\cal B}$ that is asymptotic to $\Bbb{R}^4$ one must therefore have  $q_j \in \Bbb{Z}$, $q_j \ge 0$ and $q_0 =1$ and this has precisely one solution which may easily be seen to be flat $\Bbb{R}^4$ globally if one uses the transformation  that led to (\ref{asympmet}).

\begin{figure}
\centerline{\includegraphics[ height=3 cm] {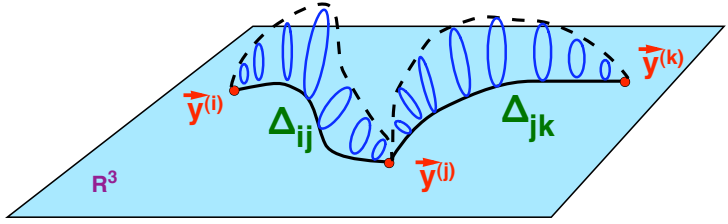}}
\caption{This figure depicts some non-trivial cycles of the Gibbons-Hawking geometry.
The behaviour of the $U(1)$ fiber is shown along curves between the geometric charges. 
Here the fibers sweep out a pair of intersecting homology spheres.}
\label{fig:NPW2}
\end{figure}

It is also easy to exhibit the homology and cohomology of the GH spaces.  Such manifolds have $N(N-1)$ non-trivial $2$-cycles, $\Delta_{ij}$, that run between the geometric charges, $q_j$.  These $2$-cycles can be defined by taking any curve, $\gamma_{ij}$, between $\vec{y}^{(i)}$ and $\vec{y}^{(j)}$ and considering the $U(1)$ fiber of (\ref{GHmetric}) along the curve.  This fiber collapses to zero at the geometric charges, and so the curve and the fiber sweep out a $2$-sphere (up to $\Bbb{Z}_{|q_j|}$ orbifolds).  See Figure \ref{fig:NPW2}.  These spheres intersect one another at the common points $\vec{y}^{(j)}$.  There are $(N-1)$ linearly independent homology two-spheres, and the set $\Delta_{i\, (i+1)}$ represents a basis.
  
%%%%%%%%%%%%%%   Figure 2   %%%%%%%%%%%%%%%
%
%
%%%%%%%%%%%%%%   End Figure 2   %%%%%%%%%%%%% 

To give the explicit harmonic $2$-forms, introduce a set of frames
\begin{equation}
\hat e^1~=~ V^{-{1\over 2}}\, (d\psi ~+~ A) \,,
\qquad \hat e^{a+1} ~=~ V^{1\over 2}\, dy^a \,, \quad a=1,2,3 \,,
\label{GHframes}
\end{equation}
and define two associated sets of two-forms:
\begin{equation}
\Omega_\pm^{(a)} ~\equiv~ \hat e^1  \wedge \hat
e^{a+1} ~\pm~ \coeff{1}{2}\, \epsilon_{abc}\,\hat e^{b+1}  \wedge
\hat e^{c+1} \,, \qquad a =1,2,3\,. 
\label{twoforms}
\end{equation}
The two-forms, $\Omega_-^{(a)}$, are anti-self-dual,  harmonic and non-normalizable  and they define the hyper-K\"ahler  structure on the base.  The forms, $\Omega_+^{(a)}$, are self-dual and can be used to construct harmonic fluxes that are dual to the $2$-cycles.  Consider  the self-dual two-form:
\begin{equation}
\Theta ~ \equiv~\sum_{a=1}^3 \, \big(\partial_a \big( V^{-1}\, H \big)\big) \,
\Omega_+^{(a)} \,.
\label{harmtwoform}
\end{equation}
Then $\Theta$ is closed (and hence co-closed and harmonic) if and only if $H$ is harmonic in $\Bbb{R}^3$, {\it i.e.}  $\nabla^2 H =0$.  For the smooth harmonic two-forms, $\Theta$ has to be regular, and so  $H/V$ must be  regular and hence
$H$ has the form:
\begin{equation}
H ~ =~  \sum_{j=1}^N \,  {h_j  \over r_j} \,.
\label{Hform}
\end{equation}
The possible  constant term in $H$   has been dropped so that $H/V$ vanishes at infinity (remember that $\varepsilon_0 =0$) and the forms are thus normalizable.  There is also a  ``gauge transformation:''
\begin{equation}
H ~ \to~ H ~+~ c\, V \,,
\label{gaugetrf}
\end{equation}
for any constant, $c$, and this leaves $\Theta$ unchanged.  Thus there are  only $N-1$ independent parameters in $H$, which matches the number of $2$-cycles.

\subsection{Ambi-polar metrics and the geometric transition in Gibbons-Hawking geometries}
 
To obtain non-trivial hyper-K\"ahler metrics that are asymptotic to $\Bbb{R}^4$ one must, of course, weaken one of the assumptions of the theorem that denies their existence.  One finds that the correct way to achieve this is to drop the insistence that the metric be Riemannian.  More precisely, the base metric should be allowed to be {\it ``ambi-polar''}, which means that it should be allowed to change its overall sign from signature $+4$ to signature $-4$.  At first sight this seems completely unphysical, but one must remember that the metric on ${\cal B}$ is an {\it auxiliary metric} and that it is only the five-dimensional metric (\ref{fivemetric}) that is required to be smooth and Lorentzian.   With an ambi-polar base metric one can still achieve a sensible physical metric  provided that the warp factors, $Z_I$, simultaneously change sign when the signature of the base metric charges sign.  Moreover, it seems to be one of the beautiful structural features of the BPS equations that guarantees that this happens.  

As I will discuss below, for the GH metrics one finds that   $Z  \sim V^{-1}$ near $V=0$ and while this certainly changes sign appropriately, it also seems to introduce singular behavior on the {\it critical surfaces}  defined by $V=0$. By another nice conspiracy of the BPS equations, all negative powers of $V$ cancel in (\ref{fivemetric}) and so the ambi-polar GH bases can give rise to smooth Lorentzian physical metrics in five dimensions  \cite{Giusto:2004kj, Bena:2005va, Berglund:2005vb}.

In a GH metric this means that one can now allow the geometric charges,   $q_j$, to be negative, but one still imposes $q_0 =1$ to get the base to be asymptotic to $\Bbb{R}^4$.   The geometric transition of a supertube or black ring is then simple to describe:  One pair-creates geometric charges underneath the ring profile.  This blows up a new pair of homology cycles in the manner depicted in Figures \ref{fig:NPW1} and  \ref{fig:NPW2} and one can replace the M5 branes by fluxes through these cycles.  Conversely, given a solution with fluxes on a GH base one can blow down a pair of equal but opposite charges by bringing them together (this typically requires taking the two geometric charges to be very large) and the result is a singular M5 brane wrapping the GH fiber.  

The new ``bubbled solutions'' thus have a large number of moduli:  one can create bubbles of arbitrary geometric charge, one can combine these charges and each geometric charge has a modulus, $\vec y^{(j)}$.    As I will discuss in the next section, the absence of CTC's  imposes some constraints on these moduli.

\section{Bubbled microstate geometries}

\subsection{Bubbled solutions on a Gibbons-Hawking base}

It is straightforward to solve the BPS equations on a GH base \cite{Gauntlett:2002nw,
Gauntlett:2004qy, Bena:2005va,Berglund:2005vb}.  The first step is to take the $\Theta^{(I)}$to be {\it
regular}, self-dual, harmonic two-forms as in (\ref{harmtwoform}) but with $H$ replaced by $K^I$ where
\begin{equation}
K^I ~=~   \sum_{j=1}^N \, {k_j^I \over r_j} \,,
\label{KIdefn}
\end{equation}
with $r_j ~\equiv~ |\vec y - \vec y^{(j)}|$.
The flux  of the two-form, $\Theta^{(I)}$, through the two-cycle $\Delta_{ij}$ is given by
\begin{equation}
\Pi_{ij}^{(I)}   ~=~       \bigg( {k^I_j   \over q_j} ~-~
 {k^I_i \over q_i} \bigg) \,, \qquad  1 \le i,j \le N \,.
 \label{fluxes}
 \end{equation}

The warp factors satisfying (\ref{BPSeqn:2}) then have the form
\begin{equation}
Z_I ~=~ \coeff{1}{2}  \, C_{IJK} \, V^{-1}\,K^J K^K  ~+~ L_I \,.\,
\label{ZIform}
\end{equation}
where the $L_I$ are three more independent harmonic functions on $\Bbb{R}^3$.
The one-form, $k$, may be written as:
\begin{equation}
k ~=~ \mu\, ( d\psi + A   ) ~+~ \omega \,,
\label{kansatz}
\end{equation}
where 
\begin{equation}
\mu ~=~ \coeff {1}{6} \, C_{IJK}\,  {K^I K^J K^K \over V^2} ~+~
{1 \over 2 \,V} \, K^I L_I ~+~  M\,,
\label{mures}
\end{equation}
and $M$ is yet another harmonic function on $\Bbb{R}^3$.  
One then finds that $\omega$ must satisfy:
\begin{equation}
\vec \nabla \times \vec \omega ~=~  V \vec \nabla M ~-~
M \vec \nabla V ~+~   \coeff{1}{2}\, (K^I  \vec\nabla L_I - L_I \vec
\nabla K^I )\,.
\label{omegeqn}
\end{equation}
The integrability condition for this equation is simply the fact
that the divergence of both sides vanish, which is true because
$K^I, L_I, M$ and $V$ are harmonic.  

Regularity requires that functions $Z_I$ and $\mu$ have no singular sources.
This means that they must have the same form as $V$:
\begin{equation}
L^I ~=~ \ell^I_0 ~+~  \sum_{j=1}^N \, {\ell_j^I \over r_j} \,, \quad
M ~=~ m_0 ~+~  \sum_{j=1}^N \, {m_j \over r_j} \,.
\label{LMexp}
\end{equation}
Moreover, regularity at $r_j =0$ means that we must take 
\begin{eqnarray}
\ell_j^I  &~=~& -  \coeff{1}{2}\,  C_{IJK} \,
{ k_j^J \, k_j^K  \over q_j} \,,  \qquad j=1,\dots, N \,;\\
m_j &~=~&   \coeff {1}{12}\,C_{IJK} {k_j^I \, k_j^J \, k_j^K \over q_j^2}  ~=~
\coeff{1}{2}\,  {k_j^1 \, k_j^2 \, k_j^3 \over q_j^2} \,,  \qquad j=1,\dots, N \,.
\label{lmchoice}
\end{eqnarray}
Finally, $\mu$ must vanish at infinity and we normalized the warp factors so that $Z_I \to 1$ at infinity, which means we must take:
\begin{equation}
\ell_0^I =1\,, \qquad 
 m_0  = -\coeff{1}{2}\, q_0^{-1} \, \sum_{j=1}^N\, \sum_{I=1}^3 k_j^I \,.
\label{fiveDsol}
\end{equation}
It is straightforward to determine $\omega$ for this solution\cite{Bena:2005va,Berglund:2005vb} but I will not need it here.  The beauty of the GH geometry is that the complete solution can be  obtained simply and explicitly. 

One can also explicitly verify that the metric and gauge fields are completely regular and the metric Lorentzian in the neighborhood of the critical surfaces $V=0$.

\subsection{The Bubble Equations}

While one has solved the BPS equations, there is still the important condition that the solutions are required to have no CTC's.  The first, and most obvious danger is the region near the geometric charges where $r_j \to 0$.  One can see form (\ref{fivemetric}),   (\ref{GHmetric}) and (\ref{kansatz}) that there will be no  CTC's in the neighborhood of the geometric charges if and only if the function, $\mu$, vanishes as $r_j \to 0$.  This leads to the {\it Bubble Equations}:
\begin{equation}
 \sum_{{\scriptstyle j=1} \atop {\scriptstyle j \ne i}}^N \,
\,  \Pi^{(1)}_{ij} \,   \Pi ^{(2)}_{ij} \,  \Pi ^{(3)}_{ij} \   {q_i \, q_j  \over r_{ij} } ~=~
-2\, \Big(m_0 \, q_i ~+~  \coeff{1}{2} \sum_{I=1}^3  k^I_i \Big) \,,
\label{BubbleEqns}
\end{equation}
where $r_{ij} \equiv |\vec y^{(i)} -\vec y^{(j)}|$.   Another danger is that  there might be Dirac-Misner strings in $\omega$, but one can prove\cite{Bena:2005va, Berglund:2005vb} that the absence of such strings in the metric is equivalent to $\mu \to 0$ as $r_j \to 0$ and so is solved by the bubble equations.

There is a simple, intuitive meaning to these equations.  Geometric charges, $q_i$ and $q_j$, of opposite sign may be thought of as D6 branes of opposite charge and so they tend to attract.  On the other hand, the fluxes through the bubbles tend to cause them to expand and so there are (families of) equilibrium configurations where these forces balance and constrain the scales, $r_{ij}$, of the bubbles in terms of the fluxes. Also note that the left-hand side involves the product of all three fluxes and so bubbling solutions are going to generically require three $U(1)$ gauge fields if the bubbles are to attain a finite size.  This is the bubbled analog of three charges being necessary for a macroscopic horizon.

Note that there are $N$ bubble equations but that the sum of them is trivially zero, and so they represent $N-1$ constraints 
on the $3(N-1)$-dimensional parameter space of the $\vec y^{(j)}$ and the $4N-1$ quantized parameters, $q_j$ and $k_j^I$  \footnote{The subtractions in $4N-1$ and  $3(N-1)$  arise because of  the constraint, $q_0 =1$, and because the center of mass of the $\vec y^ {(j)}$ is physically irrelevant.}. If one fixes the charges, $Q_I$, $J_1$ and $J_2$,  at infinity then there are five more constraints on the parameters. There is thus a huge moduli space of the solutions.   There are however, limits.  With fixed asymptotic charges one cannot have arbitrarily many bubbles because one must have at least one quantum of each flux to hold up a given bubble and the number of flux quanta is limited by the asymptotic charge. 
 
While it is easy to demonstrate that the  solutions described here are free of pathology near $V=0$ and free of CTC's near the geometric charges, one must still check that the solutions are globally free of CTC's.  This is not always true, but the counterexamples always seem to involve clusters of bubbles that have a net negative charge and have CTC's and pathological regions for the same mundane reason that a ``supersymmetric solution''   with two BPS black holes, one with charge $+Q$ and one with charge $-Q$, have pathological regions in between.  As yet we have no general theorems to cover this, but we do have a  very large number of examples that have undergone extensive numerical tests to show that not only are there no CTC's but the solutions are {\it stably causal} with a globally defined time function.  In particular, there are no horizons in these microstate geometries.

\subsection{Asymptotic charges}

It is easy to read off the electric charges of the bubbled configurations from the asymptotic behavior of the $Z_I$.  One finds:
\begin{equation}
Q_I ~=~ -2 \, C_{IJK} \, \sum_{j=1}^N \, q_j^{-1} \,
\tilde  k^J_j \, \tilde  k^K_j\,,
\label{QIchg}
\end{equation}
where
\begin{equation}
\tilde  k^I_j ~\equiv~ k^I_j ~-~    q_j\, N  \,  k_0^I  \,,
\qquad {\rm and} \qquad k_0^I ~\equiv~{1 \over N} \, \sum_{j=1}^N k_j^I\,.
\label{ktilde}
\end{equation}
Note that $\tilde  k^I_j$ is gauge invariant under (\ref{gaugetrf}) with $H$ replaced by $K^I$.

One can obtain the angular momenta from the behavior of the one-form, $k$, at infinity.  The simplest is $J_R$, which is conjugate to translations  along the GH fiber, $\psi$:
\begin{equation}
J_R ~\equiv~ J_1 + J_2 ~=~ \coeff{4}{3}\, \, C_{IJK} \, \sum_{j=1}^N q_j^{-2} \,
\tilde  k^I_j \, \tilde  k^J_j \,  \tilde  k^K_j  \,.
\label{Jright}
\end{equation}
The other angular momentum, in the $\Bbb{R}^3$ base, depends upon the geometric layout.  One first 
defines the dipoles
\begin{equation}
\vec D_j ~\equiv~  \, \sum_I  \, \tilde k_j^I \, \vec y^{(j)} \,,
\qquad \vec D ~\equiv~ \sum_{j=1}^N \, \vec D_j \,.
\label{dipoles}
\end{equation}
and then one finds
\begin{equation}
 J_L ~\equiv~ J_1 - J_2 ~=~ 8 \,\big| \vec D\big|  \,.
\label{Jleft}
\end{equation}
While there is a modulus sign around $\vec D$ in (\ref{Jleft}), one
should note that it does have a meaningful orientation.

 \section{Scaling geometries, fluctuating geometries and entropy enhancement}

It is now evident that there are a vast number of bubbled, smooth horizonless geometries that have the same asymptotic structure at infinity as a given black hole or black ring.   That is, there are a vast number of microstate geometries for black holes and black rings.   Indeed, it is relatively simple to construct many bubbled microstate geometries by sprinkling roughly equal fluxes onto any number of bubbles but one finds that this tends to produce microstate geometries corresponding to  maximally spinning black holes with vanishing horizon area \cite{Bena:2006is}.   The construction of microstate geometries for black holes with macroscopic horizons was discovered through the study of mergers \cite{Bena:2005zy} and the first microstate geometries corresponding to black holes and black rings with macroscopic horizons were then discovered \cite{Bena:2006kb, Bena:2007qc}.  The essential new element in this construction was the importance of {\it deep} or {\it scaling} microstate geometries.
 
\subsection{Scaling geometries}
\label{ScalingGeoms}

To obtain a scaling solution one focusses on a subset, ${\cal S}$, of the geometric charges and keeps  the fluxes on this subset large, so that the associated electric charges are large, but tunes the fluxes so that the bubble equations allow the geometric  charges  approach one another arbitrarily closely  in the geometry of $\Bbb{R}^3$.   That is, one looks for solutions to the bubble equations with large, and typically non-parallel,  flux parameters for which one also has $r_{ij} \to 0$ for $i,j \in {\cal S}$.  

While this appears to be a rather singular limit in terms of the (unphysical) $\Bbb{R}^3$ geometry, in the physical geometry, with all the warp factors,   the points in ${\cal S}$   remain essentially at a fixed distance from each other and descend  a long black-hole-like throat.  In the intermediate region, between the cluster defined by ${\cal S}$ and infinity, one has $Z_I \sim {\hat Q_I \over 4 \, r}$,  where I have taken ${\cal S}$ to be centered at $r =0$ and the $\hat Q_I$ are the electric charges associated with ${\cal S}$.  The fact that the warp factors behave as  $Z_I \sim r^{-1}$ leads to the macroscopic physical size of the cluster and to an $AdS$  throat precisely like that of a BPS  black hole.  

Thus, in this intermediate regime,  the scaling microstate geometry  is  almost identical to the throat of a black hole or black ring (depending upon the total geometric charge in ${\cal S}$).  This class of microstate geometries therefore appears to be exactly like a spherically symmetric black hole or black ring until one gets extremely close to where the horizon might have been.   In this region the microstate geometry caps off in a cluster, or foam, of bubbles whose physical size is contained by a surface whose area matches that of the horizon of the corresponding black object.   In this way one gets microstate geometries for black holes and black rings with macroscopic horizons.

One can construct many examples of smooth, bubbled microstate geometries of both black holes and black rings and one can check all the details of the foregoing picture.  Indeed, one can explicitly check that these microstate geometries have charges $Q_I$, $J_L$ and $J_R$ for which the corresponding black hole or black ring does indeed have a macroscopic horizon\cite{Bena:2006kb, Bena:2007qc}.   Moreover,   these examples can be arranged to have large bubbles, and hence small curvatures compared to the Planck scale.  Thus these microstate geometries lie well within the range of validity of the supergravity approximation. 

One other very important aspect of these solutions is that they now exhibit a long $AdS$ throat and so one can use the techniques of holography and argue that every microstate geometry, since it is smooth and horizonless, must correspond to a unique state in the dual holographic field theory.  In particular, by going to the D1-D5-P duality frame one can match geometries to states in the strongly coupled D1-D5-P  field theory on the boundary.  The  states of this field theory were originally counted at weak coupling and this led to the match with the black-hole entropy but now one can use the AdS/CFT correspondence to match strongly coupled states to non-trivial microstate geometries.     One is therefore not simply limited to making purely semi-classical arguments about geometric configurations:  One can use holography to put conceptual and computational flesh on the whole idea.

Given these solutions, the obvious question is whether there might be enough of them to account for the entropy semi-classically. There is an obvious topological entropy arising from partitioning the charges in terms of flux amongst bubbles and then there is an entropy associated with quantizing the moduli space of positions of the geometric charges.  These entropies can easily be estimated \cite{Bena:2006is} for non-scaling solutions and one finds that they give $S_{top} \sim Q^{1/4}$  and $S_{moduli} \sim Q^{1/2}$ respectively, which is far short of the needed black-hole entropy:
\begin{equation}
S~=~    2 \pi\,  \sqrt{Q_1 Q_2 Q_3 ~-~ J_R^2} ~\sim~ Q^{3 \over 2} \,.
 \label{BHent}
\end{equation}
So we have a vast number of solutions, but nowhere near enough.

On the other hand, one should look at scaling solutions and one should study small fluctuations around such backgrounds \cite{Bena:2006kb}.  A rather simple estimate shows that these excitations have energies that match those of states in the typical sector\footnote{By typical, I mean the sector that gives the largest contribution to the black-hole entropy.} of the dual conformal field theory on the boundary.  It therefore seems likely that the scaling, bubbled microstate geometries are representatives of states that lie in precisely the physical sector of the theory that we wish to study. Indeed, it seems that we have geometries that describe a relatively sparsely distributed sample of the microstates of most interest.  The obvious question is whether we can get more such geometries and thereby sample the microstates of the system more completely.  Indeed, it futher suggests that one should look at fluctuating geometries around the scaling bubbled solutions presented above.    I will discusss this below.

Before proceeding with this, it is already worth noting that the microstate geometries already raise a very interesting set of questions.  First there is the mapping out of the holographic correspondence for these geometries and understanding precisely which microstates they represent  \cite{Kanitscheider:2006zf}.  Then there is the quantization of the moduli space of the deep or scaling microstate geometries  \cite{deBoer:2008zn}.   There is also the interesting question about how the work I am discussing here is related to the work on attractor flows, which also singles out scaling geometries as the sector of the theory that provides a number of states that are sufficient to account for the field theory entropy  \cite{Denef:2007vg, Denef:2007yt}.    I will return to this last issue in the conclusions.  

Finally, it is also worth recalling the simple physical point that, given the failure of black hole uniqueness and existence of many smooth horizonless geometries, one must revisit the issue of what it means physically to select singular solutions with a horizon and, once again, ask the question:  Is the spherically symmetric solution merely an approximate or effective solution that simply gives some kind of average effect of all the microstate geometries?

\subsection{Fluctuating geometries and entropy enhancement}

Two-charge supertubes have played a major role in the study of the microstate geometries for two-charge black holes.  This is because perturbative supertubes \cite{Mateos:2001qs} are relatively easy to quantize and one can also find the exact corresponding supergravity solutions with arbitrary classical profiles and, in the D1-D5 duality frame, these supergravity solutions are completely regular\cite{Lunin:2001jy,Lunin:2002iz}.  Thus we may think of these solutions as two-charge microstate geometries.  The arbitrary shapes of supertubes can lead to a lot of entropy but  their naive quantization cannot  hope to account for the entropy of a black hole with a non-trivial, macroscopic horizon.  Indeed,  since supertubes only carry two charges, their entropy scales like \cite{Palmer:2004gu,Bak:2004kz,Rychkov:2005ji}:
\begin{equation}
S  ~\sim~    \sqrt{Q_1 \, Q_2 } ~\sim~  Q  \,.
 \label{Stube}
\end{equation}
In addition there is the  concern that the  two-charge black hole has a Planck scale horizon and so the typical microstate geometries must lie at the edge of the validity of the supergravity approximation.

On the other hand it is natural to ask whether such fluctuating geometries can be incorporated into the three-charge bubbled geometries and generalized or extended  to give even more entropy  \cite{Bena:2008wt, Bena:2008nh}.   At the very least, (\ref{Stube})  represents more entropy than that coming from the topological entropy and the  GH moduli space. 

To describe the regular supertube geometries one must pass to the D1-D5-P duality frame and lift the bubbled solutions to six-dimensional geometries arising from a $T^4$ compactification of IIB supergravity.  Having done this one can then incorporate round supertubes into the GH base by a spectral flow transformation that arises from a simple global coordinate change that mixes $U(1)$ fibers of the six-dimensional geometry \cite{Bena:2008wt}.  One can then recast fluctuating supertubes in terms of fluctuating bubbled geometries.  This generates even more general classes of microstate geometries because one is now considering non-trivial fluctuations in more than five dimensions.  

At present the complete supergravity solution for a fluctuating supertube in a scaling microstate geometry has not been constructed.  However, one can perform a probe calculation using the Dirac-Born-Infeld action in such a microstate geometry.  Furthermore, because one knows that the corresponding full supergravity solution must be completely regular, one has confidence that such a probe calculation will capture essential features of the fully back-reacted fluctuating solution. 

The results of this calculation are rather remarkable\cite{Bena:2008nh} and one finds that the ability of a supertube to store entropy,  (\ref{Stube}), is governed, not by its charges measured at infinity, but by its {\it effective charges}, $Q_I^{eff}$.  To be more precise, the effective charges are defined by the near-tube divergences\footnote{ The $Z_I$ are  finite and regular in the five-dimensional pure bubbled geometries, but incorporating supertubes allows some of the $Z_I$ to diverge near the supertubes.  These singularities do not affect the regularity of the geometry overall, but the coeefficient of the divergence defines the local effective charge.} of the $Z_I$ and not by the behavior of the $Z_I$ at infinity. From the form of the $Z_I$, (\ref{ZIform}), one sees that these effective charges are a combination of the supertube charges and the interaction between the supertube magnetic dipole moment and the background magnetic dipole fields.   These effective charges can become huge in scaling bubbled geometries, particularly near the critical surfaces.  

Thus embedding fluctuating supertubes into a deep scaling solution can enable the supertube to store vastly more entropy than the corresponding object in ordinary space time.  This phenomenon is called {\it entropy enhancement}.  It is therefore possible that fluctuating bubbled geometries might provide a sufficiently good sample of microstate to enable a semi-classical account of black-hole entropy.  This is currently under very active study.

Another very encouraging sign is the remarkable convergence of thought between this work on microstate geometries and the work related to black-hole deconstruction  \cite{Denef:2007vg , Denef:2007yt, Gimon:2007mha}.  This work has also identified deep, scaling configurations and critical surfaces as being essential to the description of the macroscopic black-hole  entropy.   In this approach, the string coupling is of ``intermediate magnitude''  where the gravitational back-reaction of some of the branes is neglected.   One can then, for example,   use index theory in supersymmetric quantum mechanics to count the (index of) ground states.  The results show that for precisely the scaling solutions, and no others, the number of states grows at a sufficient rate to account for the macroscopic black-hole entropy.  The difference between the approach described here and that of deconstruction is that here the focus is on obtaining microstate  ({\it i.e. smooth, horizonless}) geometries.  Typically in deconstruction, at least one class of brane is treated perturbatively and, unlike the approach above, it is unclear what these perturbative branes will become, and how the geometry will change, once the back-reaction is incorporated.  Thus, while the incorporation of the back-reaction of some of the branes refines the physical picture compared to the original perturbative state counting, the naive strong-coupling extrapolation of these microstate configurations will probably not be reliable  when the classical black hole exists.   This contrasts with the regular behavior of the geometry of the two-charge supertube once its back-reaction is included.  On the other hand, while the approach and emphasis are different, both bodies of work point to exactly the same geometric structures as being the key ingredients in describing the microstates of black holes.

\section{Some final comments}

It is evident that black-hole uniqueness  is violated on a remarkable scale in string theory and M-theory.   Microstate geometries provide rich and interesting families of smooth, horizonless solutions that have the same asymptotics at infinity as a supersymmetric black hole or black ring.  While this is interesting in its own right, it  is also possible that there might be enough microstate geometries to account for the classical black-hole entropy.   It is also equally evident that one will not be able to use the supergravity approximation to describe {\it every} black hole microstate but to use this as a reason to downplay the importance of the semi-classical approach entirely misses some important physical insights.

The real issue when it comes to semi-classical counting of states and determining entropy is one of {\it sampling}.  It is simply not necessary to count anywhere near all the states of the system, one only has to account for a suitably dense subset.  Indeed, for the black-hole entropy, the burden of being of ``suitably dense'' is  very mild.  Note that in order to get (\ref{BHent}) one only need count states to an accuracy of:
\begin{equation}
1  \quad {\rm in} \quad  e^{Q^\alpha}  \,,
 \label{sampling}
\end{equation}
for $\alpha < {3 \over 2}$.  For a large black hole, this sampling is {\it incredibly sparse.} When one thinks of the problem in these terms, it is very conceivable that semi-classical computations could find enough states to account for the entropy of the system.  Indeed, it is worth remembering that in the simple, classical description of the entropy of an ideal gas, all one needs to know is that the system is quantized but the scale of that quantization is sub-leading.  For example, the entropy of a monatomic ideal gas is given by:
\begin{equation}
S ~=~ N\,k \, \Big[ \log\Big({V \over N} \Big) ~+~ \frac{3}{2}\, \log T  ~+~ 
\frac{3}{2}\, \log\Big({2 \pi m k \over \hbar^2} \Big)  ~+~  \frac{5}{2} \Big] \,.
 \label{IdealGas}
\end{equation}
The value of $\hbar$ gives a sub-leading correction to this entropy, and so if one uses a crude classical approximation to quantization one gets the correct leading thermodynamic behavior.   One might therefore expect the supergravity approximation of black hole microstates to provide, at least, the leading order thermodynamic behavior.   

The picture of a monatomic ideal gas as a set of elastic spheres bouncing around in a box is a very valuable theoretical tool even though it is a crude classical approximation. In reality, we know that this ``billiard ball'' picture is a modest  subset of rather special coherent states of the underlying quantum system, but it is enough to get some very useful physical understanding.  It is thus not unreasonable to hope that the same may be true of microstate geometries.  Indeed, as I discussed in Section \ref{ScalingGeoms}, the bubbled microstate geometries appear to be sampling precisely the correct typical sector of the dual CFT.  

While on the issue of microstates and sampling, it is also worth remembering that the other approaches to counting states, like the OSV conjecture\cite{Ooguri:2004zv}, also rely upon sampling and do not necessarily count all the states. The counting is done using a topological index and for every state that contributes to the index there are potentially many that do not.  Indeed, if the black hole is ``too supersymmetric,'' as it is when one compactifies the IIB theory on $T^6$ or $K3 \times T^2$, then the topological index counts essentially none of the states.  However, the OSV conjecture has met with spectacular success for compactifications on generic Calabi-Yau manifolds,  and there is an incredible degree of matching even at sub-leading orders.   On the other hand, there are mismatches and the topological partition function needs to be refined to capture missing BPS states \cite{Dabholkar:2005dt, Denef:2007vg}. Since we are counting BPS states of a BPS object, it seems likely that the topological index can eventually be refined far enough to produce a complete count, but at present there are still discrepancies.

With the success of the OSV conjecture and the much deeper understanding of BPS microstate geometries, it seems that we have gained a much deeper understanding of the quantum structure of BPS black holes.  It is thus natural to ask what parts of this description,  if any, survive in the description of microstates of non-supersymmetric, non-extremal black holes.  From the perspective of microstate geometries rather little is known but there are now examples of a non-supersymmetric microstate geometries \cite{Jejjala:2005yu, Giusto:2007tt}.   One could also study families of near-BPS states.  In particular, given the restrictions imposed by the bubble equations on scaling microstate geometries one still has free moduli but their ranges should be compact.   One can therefore look for non-BPS solutions that involve slow motion on this moduli space.  Such solutions will almost certainly exist, provided that there is no unstable acceleration, and will yield non-BPS solutions that preserve many of the desirable topological features of the solutions presented here.

It remains to be seen whether anything can be said about a Schwarzschild or Kerr black hole, but one of the most interesting ideas to emerge out of the geometric approach to black-hole microstates is that near a black hole, string theory and D-branes may generate significant numbers of large collective excitations whose  scales are much larger than the string scale or Planck scale.  Such excitations will almost certainly play a role in understanding what happens inside and in the neighborhood of a black hole and perhaps ultimately in the observational testing of string theory.

\bigskip 
\section*{Acknowledgements}
\smallskip
I would like to thank the Prof. Eguchi and the Organizers of this superb conference for the opportunity to attend and speak.   I am also very grateful to my collaborators, particularly Iosif Bena, without whom this work would simply not have been possible.  Finally, this work was supported in part by funds provided by the DOE under grant DE-FG03-84ER-40168.

%\appendix
%\section{First Appendix} %Empty argument \section{} yields `Appendix'. 
%
%\section{Second Appendix}

\end{document}